\documentclass[lnbip]{svmultln}

\usepackage{graphicx}
\usepackage{booktabs}
\usepackage{paralist}
\usepackage{todonotes}
\usepackage{makeidx} 
\usepackage{multirow}
\usepackage{pdflscape}
\newcommand*\rot{\rotatebox{90}}

\begin{document}
\mainmatter              
\title{On Evidence-based Risk Management in Requirements Engineering}
\titlerunning{On Evidence-based RE Risk Management}  
%
\author{Daniel M\'{e}ndez Fern\'{a}ndez \inst{1} \and Michaela Tie{\ss}ler\inst{1} \and Marcos Kalinowski\inst{2} \and Michael Felderer\inst{3} \and Marco Kuhrmann\inst{4} }
\authorrunning{D. M\'{e}ndez Fern\'{a}ndez et al.}   

\tocauthor{Daniel M\'{e}ndez Fern\'{a}ndez, Michaela Tie{\ss}ler, Marcos Kalinowski, Michael Felderer, Marco Kuhrmann}
\institute{Technical University of Munich, Germany\\
\email{Daniel.Mendez | Michaela.Tiessler@tum.de} 
\and
Pontifical Catholic University of Rio de Janeiro, Brazil\\
\email{kalinowski@inf.puc-rio.br}
\and
University of Innsbruck, Austria\\
\email{michael.felderer@uibk.ac.at}
\and
Clausthal University of Technology, Germany\\
\email{kuhrmann@acm.org}
}

\maketitle              

\begin{abstract}

\textbf{Background:} The sensitivity of Requirements Engineering (RE) to the context makes it difficult to efficiently control problems therein, thus, hampering an effective risk management devoted to allow for early corrective or even preventive measures.
 
\textbf{Problem:} There is still little empirical knowledge about context-specific RE phenomena which would be necessary for an effective context-sensitive risk management in RE.

\textbf{Goal:} We propose and validate an evidence-based approach to assess risks in RE using cross-company data about problems, causes and effects. 

\textbf{Research Method:} We use survey data from 228 companies and build a probabilistic network that supports the forecast of context-specific RE phenomena. We implement this approach using spreadsheets to support a light-weight risk assessment. 

\textbf{Results:} Our results from an initial validation in 6 companies strengthen our confidence that the approach increases the awareness for individual risk factors in RE, and the feedback further allows for disseminating our approach into practice.

\end{abstract}
\section{Introduction}

Requirements Engineering (RE) has received much attention in research and practice due to its importance to software project success. Precise and consistent requirements directly contribute to appropriateness and cost-effectiveness in the development of a system~\cite{NE00} whereby RE is a critical determinant of productivity and software (process) quality~\cite{DC06}. Yet, RE remains an inherently complex discipline due to the various individual socio-economic and process-related influences in the industrial environments. In contrast to many other disciplines, RE is largely characterised by the particularities of the application domain, and it is largely driven by uncertainty and also by human factors. All this influences the way of working in individual contexts including the choice of methods, approaches, and tools in RE. In fact, what might work very well in one project setting might be completely alien to the needs and the culture of the next~\cite{MW2013}. The lack of possibilities to standardise the discipline with universal solutions renders it consequently difficult to efficiently control and optimise the quality of RE as an appropriate means to an end. 

In consequence, the individual particularities of the contexts hamper an effective risk management devoted to allow for early corrective or even preventive measures. Hence, it is not surprising that despite the importance of excellence in RE, we can still observe various problems in industrial environments rooted all in insufficient RE~\cite{fernandez2015naming, MW+16}. Requirements Engineering risk factors which might be typical for one project setting, might not be valid for others; for instance, while one project might be characterised by frequent changes in the requirements, the requirements in another project might be stable. Another factor to be considered is the currently weak state of empirical evidence in RE~\cite{fernandez2015naming, MW+16}. The lack of proper empirical figures that would demonstrate what phenomena occur in practical settings, what problems practitioners face, and what success factors we can infer for the various contexts makes it impossible for available risk management approaches to consider context-sensitive risk factors and related corrective and even preventive measures for RE.

To efficiently consider the particularities of RE in risk management approaches, we postulate that we first need knowledge about the current state of practice in industrial environments and problems faced therein. Motivated by this situation, we initiated the \emph{Naming the Pain in Requirements Engineering} (NaPiRE) initiative (see Sect.~\ref{sec:relatedwork}). NaPiRE constitutes a globally distributed family of bi-yearly replicated surveys on the industrial status quo and problems in RE. This allows to steer research in RE in a problem-driven manner and the establishment of proper solutions based on a better understanding of the needs of the various industrial contexts.

In this paper, we contribute the first steps towards a holistic evidence-based RE risk management approach that explicitly takes into account the particularities in RE as described above. We concentrate on assessing risk factors in RE. In particular, we use the cross-company data from NaPiRE to build a holistic model of context-sensitive problems, causes, and effects. We use this knowledge to implement a probabilistic network that allows to calculate the posterior probability of certain risk factors based on knowledge about the current situation. We use this to propose a first evidence-based RE risk assessment approach which we validate in 6 companies. We conclude by outlining current work on integrating guidelines to mitigate and correct problems in RE to make the consequential next step from a RE risk assessment approach to a holistic, evidence-based RE risk management approach. 

Our contribution and the feedback by the practitioners supports us and other researchers already in proposing further solutions to requirements engineering that need to rely on cross-company data. Further, our proposed risk assessment approach already allows practitioners to reflect on their current situation by increasing their awareness for potential problems in RE.

The remainder of this paper is organised as follows: In Sect.~\ref{sec:relatedwork} we introduce the background and related work. In Sect.~\ref{sec:researchmethodology}, we elaborate on the research design. In Sect.~\ref{sec:EbRE}, we describe our overall risk assessment approach and in Sect.~\ref{sec:validation}, we report on a validation conducted in 6 companies, before concluding our paper in Sect.~\ref{sec:conclusion}.

\section{Background and Related Work}
\label{sec:relatedwork}

Software risk management constitutes means to efficiently assess and control risk factors affecting the overall software development activities~\cite{Boehm91} (planned ones and deviations) and is often associated with ``project management for adults'' as baptised by deMarco et al~\cite{deMarco+2003}. The importance of risk management for software engineering in general and requirements engineering in particular has been addressed in several risk management approaches tailored to software engineering processes~\cite{Pfleeger2000RiskyBusiness}. Already the spiral model of Boehm~\cite{Boehm1988SpiralModel} explicitly includes risk management within software development. The Riskit method~\cite{Kontio1999RiskManagementInSoftwareDevelopment} provides a sound theoretical foundation of risk management with a focus on the qualitative understanding of risks before their possible quantification. Karolak~\cite{Karolak1995SoftwareEngineeringRiskManagement} proposes a risk management process for software engineering that contains the activities risk identification, risk strategy, risk assessment, risk mitigation, and risk prediction. With ISO 31000~\cite{iso200931000}, which was released in 2009, there is even a family of standards for risk management available that can also be instantiated in software engineering and its subareas like testing, where ISO 31000 has already been applied in the context of risk-based testing~\cite{felderer2014taxonomy}, or requirements engineering. 

A recent study on industrial development practices~\cite{Kuhrmann:2017aa} shows that practitioners see the need to explicitly include traditional risk management approaches into RE that tends to be done rather agile. This further strengthens our confidence in the need to tailor risk management approaches to the particularities of RE actively taking into account the volatility therein (as discussed in our introduction).

In fact, most work on risk management in the context of requirements engineering focuses on identifying risks in a bottom-up approach and analysing risks during the requirements engineering process. For instance, Asnar et al.~\cite{asnar2011goal} provide a goal-driven approach for risk assessment in requirements engineering, and Haisjackl et al.~\cite{haisjackl2013riscal} provide an estimation approach and respective tool support~\cite{haisjackl2013riscal}. For risk management within requirements engineering itself, Lawrence et al.~\cite{lawrence2001top} provide a list of top risks in requirements engineering, which includes overlooking a crucial requirement, inadequate customer representation, modelling only functional requirements, not inspecting requirements, attempting to perfect requirements before beginning construction as well as representing requirements in the form of designs. However, evidence-based approaches to risk management in requirements engineering as proposed in this paper are so far not available. For such an approach, we need a proper empirical basis on requirements engineering which has been recently established by the NaPiRE (Naming the Pain in Requirements Engineering) initiative. 

The NaPiRE initiative was started in 2012 in response to the lack of proper empirical figures in RE research. The idea was to establish a broad survey investigating the status quo of RE in practices together with contemporary problems practitioners encounter. This should lead to the identification of interesting further research areas as well as success factors for RE. We created NaPiRE as a means to collaborate with researchers from all over the world to conduct the survey in different countries. This allows us to investigate RE in various cultural environments and increase the overall sample size. Furthermore, we decided to run the survey every two years so that we can cover slightly different areas over time and have the possibility to observe trends. NaPiRE aims to be open, transparent and anonymous while yielding accurate and valid results. 

At present, the NaPiRE initiative has over 50 members from more than 20 countries mostly from Europe, North-America, South-America, and Asia. There have been two runs of the survey so far. The first was the test run performed only in Germany and in the Netherlands in 2012/13. The second run was performed in 10 countries in 2014/15. All up-to-date information on NaPiRE together with links to instruments used, the data, and all publications is available on the web site \url{http://www.re-survey.org}. The first run in Germany together with the overall study design was published in~\cite{fernandez2015naming}. It already covered the spectrum of status quo and problems. Overall, we were able to get full responses from 58 companies to test a proposed theory on the status quo in RE. We also made a detailed qualitative analysis of the experienced problems and how they manifest themselves. For the second run, we have published several papers~\cite{kalinowski:seke15, mendez:softw15, kalinowski:swqd16, wagner2017requirements} concentrating on specific aspects and the data from only one or two countries and one paper~\cite{MW+16} focusing on RE problems, causes and effects based on the complete data set covering data reported by 228 companies. An analysis of the data with a focus on risk management and evidence-based risk management in RE has not been published so far.

\section{Research Design}
\label{sec:researchmethodology}

Our overall goal is to provide first steps towards a holistic evidence-based RE risk management approach that allows to steer a context-sensitive risk management based on empirical cross-company data and, thus, bridges shortcomings of currently available approaches. Our research, therefore, needs to rely on data reflecting practical problems in RE that are typical for certain context factors. The scope of validity (and relevance) of the proposed solution consequently depends on the practical contexts from which the data was obtained and where we applied our approach. 

Our research design is therefore strongly inspired by the (design science) engineering cycle as exemplary described by Wieringa et al. in~\cite{MA12}. That is, we follow a cyclic development approach which is initially based on idealised assumptions, solution design, and validation as well as evaluation in practice, before revising our assumptions. With each iteration, we can make more realistic assumptions and scale our solution proposal and its effects up to practice. However, in contrast to classical design science research aimed at practical problem solving, where we develop individual solutions to practical problems, our aim is not to solely understand the effects of such solution proposals. Instead, our aim is also to better understand the particularities of contexts and phenomena involved and to actively incorporate them in our solution. That is, we begin with understanding which problems practitioners experience in certain contexts, develop our evidence-based RE Risk Assessment approach and -- by transfer into practice -- we can use the observations to make our context assumptions more reliable and precise for the next iteration. 

\begin{figure}[!hbtp]
\centering
  \includegraphics[width=0.8\textwidth]{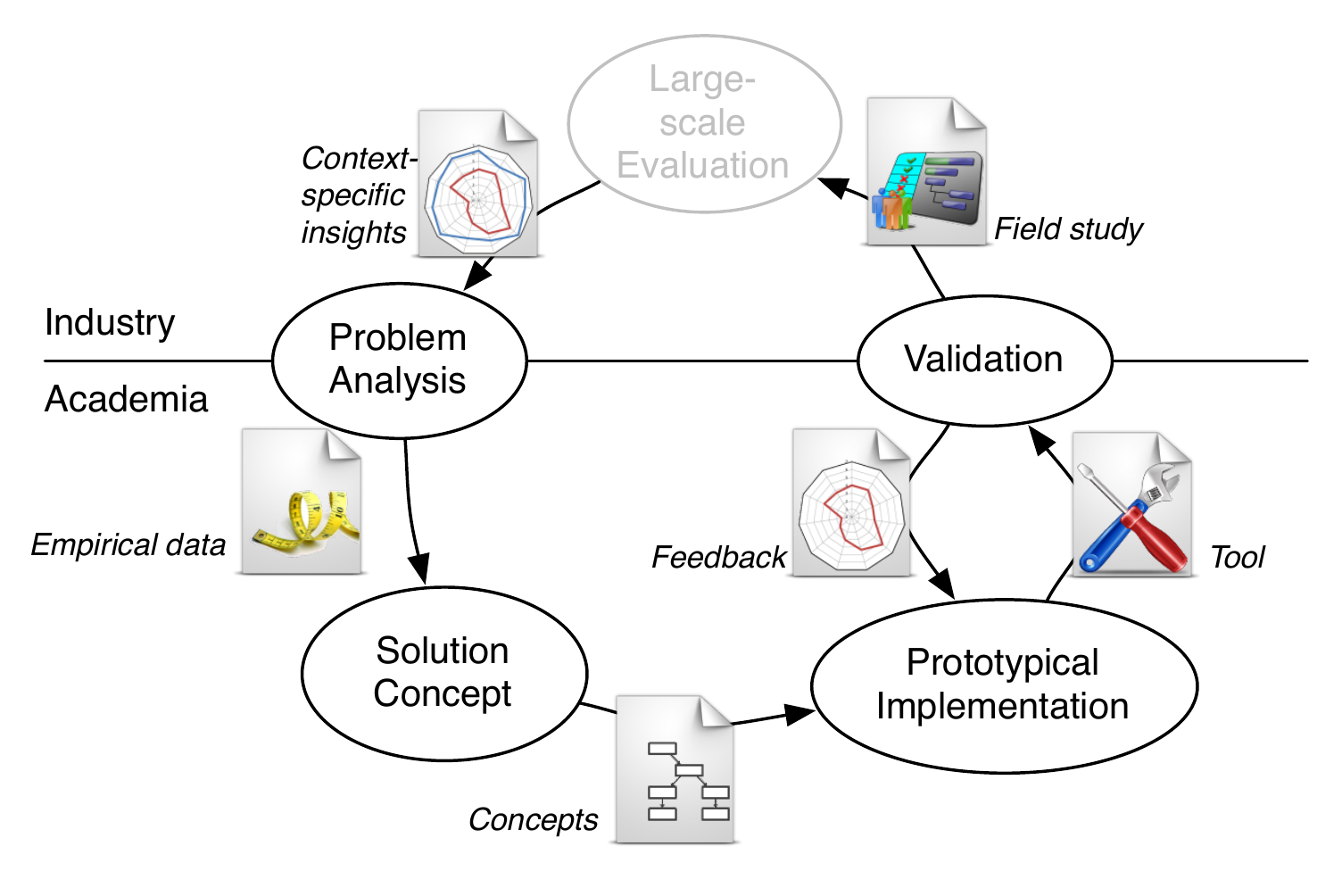}\\
  \caption{Overall research design. The last step including a field study as part of a large-scale evaluation is in scope of current activities and out of scope of this paper.}
  \label{fig:ResearchDesign}
\end{figure}

The resulting methodological approach reflects the basic notion of knowledge transfer. Our model is therefore structured in analogy to the technology transfer model as described by Gorschek et al.~\cite{GGLW06} and sketched in Fig.~\ref{fig:ResearchDesign} to visualise the cyclic nature. 

Starting with a problem analysis based on cross-company data emerging from the NaPiRE initiative, we develop a first solution proposal, which we transfer into a prototypical tool implementation. We use this prototype for a first validation with industry participants as a preparation for a large-scale evaluation. At the end of the iteration, we use the observations from the large-scale evaluation to increase our empirical data set with additional context factors before entering the next development iteration. 

In this paper, we report on the first iteration where we develop and validate our initial approach to evidence-based RE Risk Assessment and outline the next steps of a continuous evaluation and refinement towards a holistic Risk Management approach which is in scope of current work. 

In the following, we introduce the single steps to the extent necessary in context of our contribution at hands. 

\subsection{Problem Analysis}
\label{sec:ProblemAnalysis}
Our data for the initial problem analysis is largely based on the last replication of the NaPiRE initiative (see Sect.~\ref{sec:relatedwork}), including cross-company data from 228 companies located in 10 countries. There, we investigated the status quo on  the practices applied in requirements engineering as well as problems, causes, and effects as experienced and reported by practitioners from various domains. Table~\ref{tab:mostcriticalreproblems} exemplarily illustrates the most frequently cited top 10 problems as reported by the respondents starting with an accumulation of the problems and the summary of the single (top 5) ranks of the problems. 

\begin{table}[htb]
\scriptsize
\centering
\caption{Most cited top 10 RE problems as reported in~\cite{MW+16}.}
\label{tab:mostcriticalreproblems}
\begin{tabular}{p{0.6\linewidth}cccccccc}
\toprule
\textbf{RE Problem}  & \textbf{Total} & \rot{\textbf{Cause for project failure}} & \rot{\textbf{Ranked as \#1}} & \rot{\textbf{Ranked as \#2}} & \rot{\textbf{Ranked as \#3}} & \rot{\textbf{Ranked as \#4}} & \rot{\textbf{Ranked as \#5}}\\ \hline
Incomplete and / or hidden requirements	&	109	(48\%) & 43	&	34	&	25	&	23	&	17	&	10\\
Communication flaws between project team and customer	 &	93 (41\%)	&	45	&	36	&	22	&	15	&	9	&	11\\
Moving targets (changing goals, business processes and / or requirements)	&	76 (33\%)	&	39	&	23	&	16	&	13	&	12	&	12\\
Underspecified requirements that are too abstract	&	76 (33\%)	&	28	&	10	&	17	&	18	&	19	&	12\\
Time boxing / Not enough time in general	&	72 (32\%)	&	24	&	16	&	11	&	14	&	17	&	14\\ 
Communication flaws within the project team	&	62 (27\%)	&	25	&	19	&	13	&	11	&	9	&	10\\
Stakeholders with difficulties in separating requirements from known solution designs	&	56 (25\%)	&	10	&	13	&	13	&	12	&	9	&	9\\ 
Insufficient support by customer	&	45 (20\%)	&	24	&	6	&	13	&	12	&	6	&	8\\ 
Inconsistent requirements	&	44 (19\%)	&	15	&	8	&	9	&	6	&	9	&12\\
Weak access to customer needs and / or business information	&	42 (18\%)	&	16	&	7	&	10	&	8	&	8	&9\\
\bottomrule
\end{tabular}
\end{table}

We can see, for instance, that incomplete requirements dominate the list of top 10 problems in RE directly followed by communication flaws between the project team and the customer. To elaborate on risk factors and how they propagate in a project ecosystem, we need, however, knowledge going beyond single problems.

\paragraph{From Problems to Cause-Effect Chains.}
For each of the problems reported, we further analyse the data on the causes and the effects to increase our understanding about the criticality of the problems and their root causes. The latter is important to understand possibilities for corrective or even preventive measures. For instance, the main causes for communication flaws between the project team and the customer are of organisational nature, including causes such as language barriers, missing engagement by the customer, or missing direct communication with the customer. Effects of the problem include, inter alia, a poorer product quality by incorrect or missing features. A richer introduction into the problems, causes, and effects in RE can be taken from our previously published material~\cite{MW+16}.

\begin{figure}[!hbtp]
\centering
  \includegraphics[width=1\textwidth]{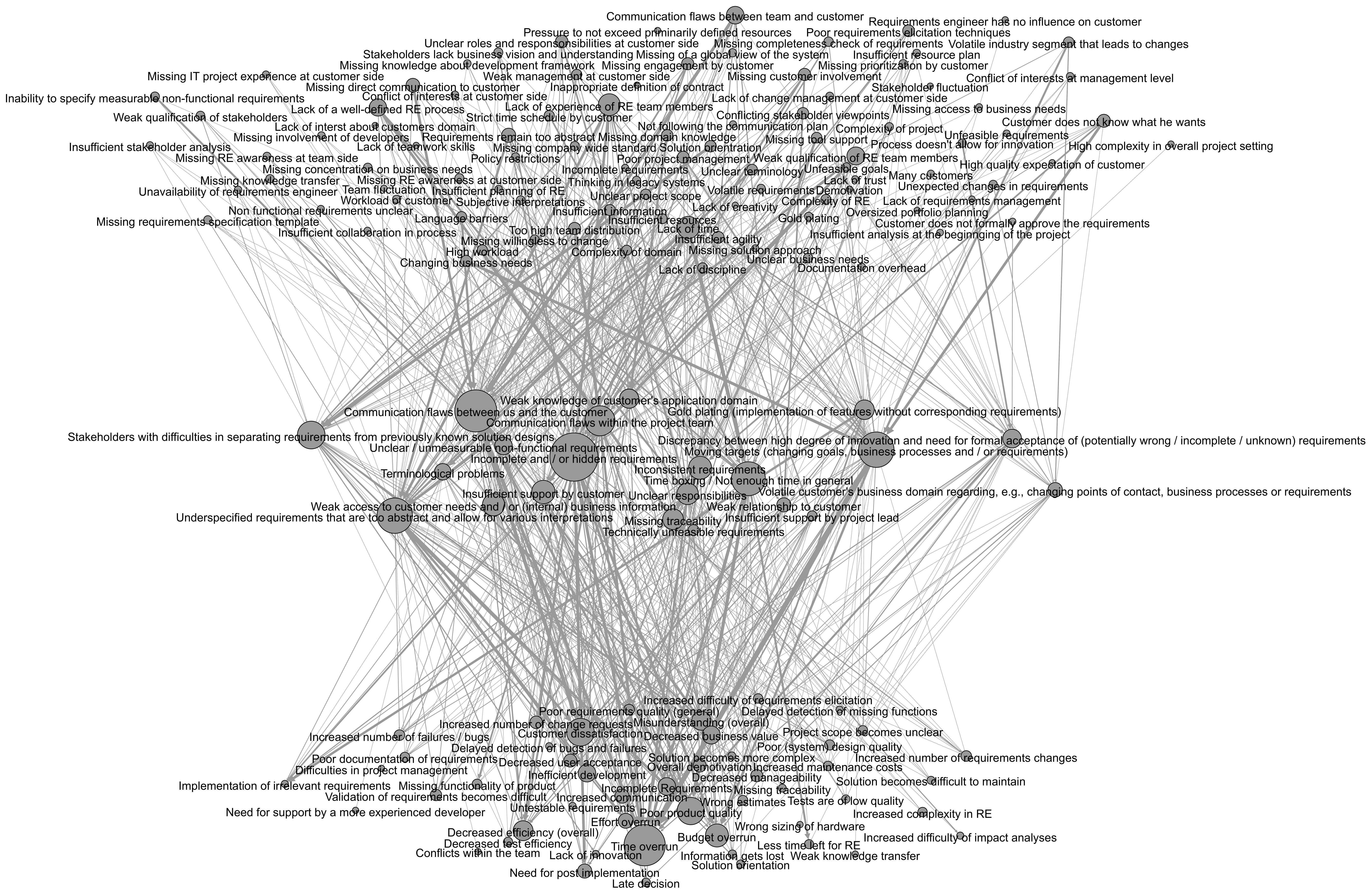}\\
  \caption{Full-scale cause-effect visualisation of the RE problems, causes and effects. Not intended to be read in detail (but we cordially invite the reader to zoom in).}
  \label{fig:CausesEffects}
\end{figure}

Figure~\ref{fig:CausesEffects} illustrates the full-scale model of the causes and the effects of the problems as distilled from the NaPiRE data set. Visualising the full-scale model serves two purposes: It shall demonstrate that (1) the model includes complex cross-dependencies and single phenomena can lead to several problems, and that (2) despite the complexity of the model, it already supports a basic understanding of the phenomena involved in software development projects and how they propagate through the different phases. Yet, such a model still does not help understanding context-sensitive conditional dependencies which renders it nearly useless for operationalisation as a risk assessment approach.
That is, for an effective evidence-based RE risk assessment, we need data that allows us to infer, at least with a certain level of confidence, those effects certain problems have based on a selection of specific context-specific conditions that characterise a project situation at any point in time.

\paragraph{From Cause-Effect Chains to Conditional Probabilistic Distributions.}

\begin{figure}[!hbtp]
\centering
  \includegraphics[width=1\textwidth]{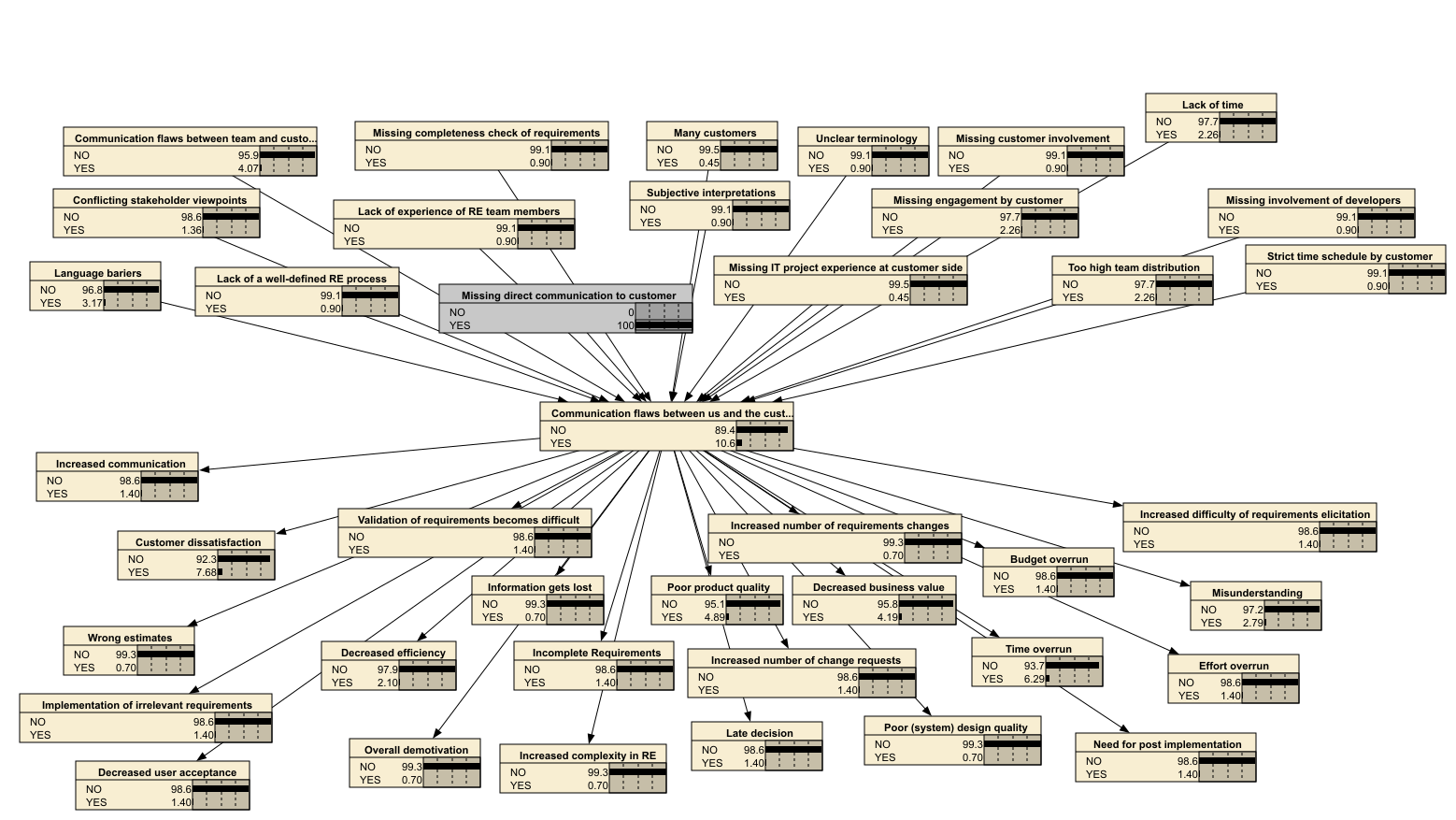}\\
  \caption{Predictive inference for effects of missing direct communication to the customer.}
  \label{fig:Netica}
\end{figure}

Finally, to lay the foundation for an evidence-based risk assessment which includes operationalisable conditional dependencies, we analyse the data by making use of Bayesian networks. We already made positive experiences in using Bayesian networks in defect causal analyses~\cite{K++17}. Here, we use Netica \footnote{\url{https://www.norsys.com/}} to implement the cause-effect information in dependency to context factors such as company size, team distribution or software process model used (agile or plan-driven). Such an implementation allows us to quantify the conditional probabilistic distributions of all phenomena involved. More precisely, it allows us to, based on the NaPiRE data used as learning set, use the Bayesian network inferences to obtain the posterior probabilities of certain phenomena to occur when specific causes are known. Figure~\ref{fig:Netica} shows such an exemplary inference for the consequences and their probabilities of the phenomena \emph{missing direct communication to the customer} (highlighted in grey). 

We use this analytically curated data as a basis for the proposal of our evidence-based risk assessment approach.

\subsection{Solution Concept and Prototypical Implementation}
Once understanding the data and its potential, we elaborate the design of our initial evidence-based RE Risk Assessment approach. This approach shall, in a first step, serve the purpose of providing means to assess risk factors in a light-weight manner based on knowledge about a current project situation -- while still offering anchor points for extensions to a holistic risk management approach.

\begin{figure}[!hbtp]
\centering
  \includegraphics[width=1\textwidth]{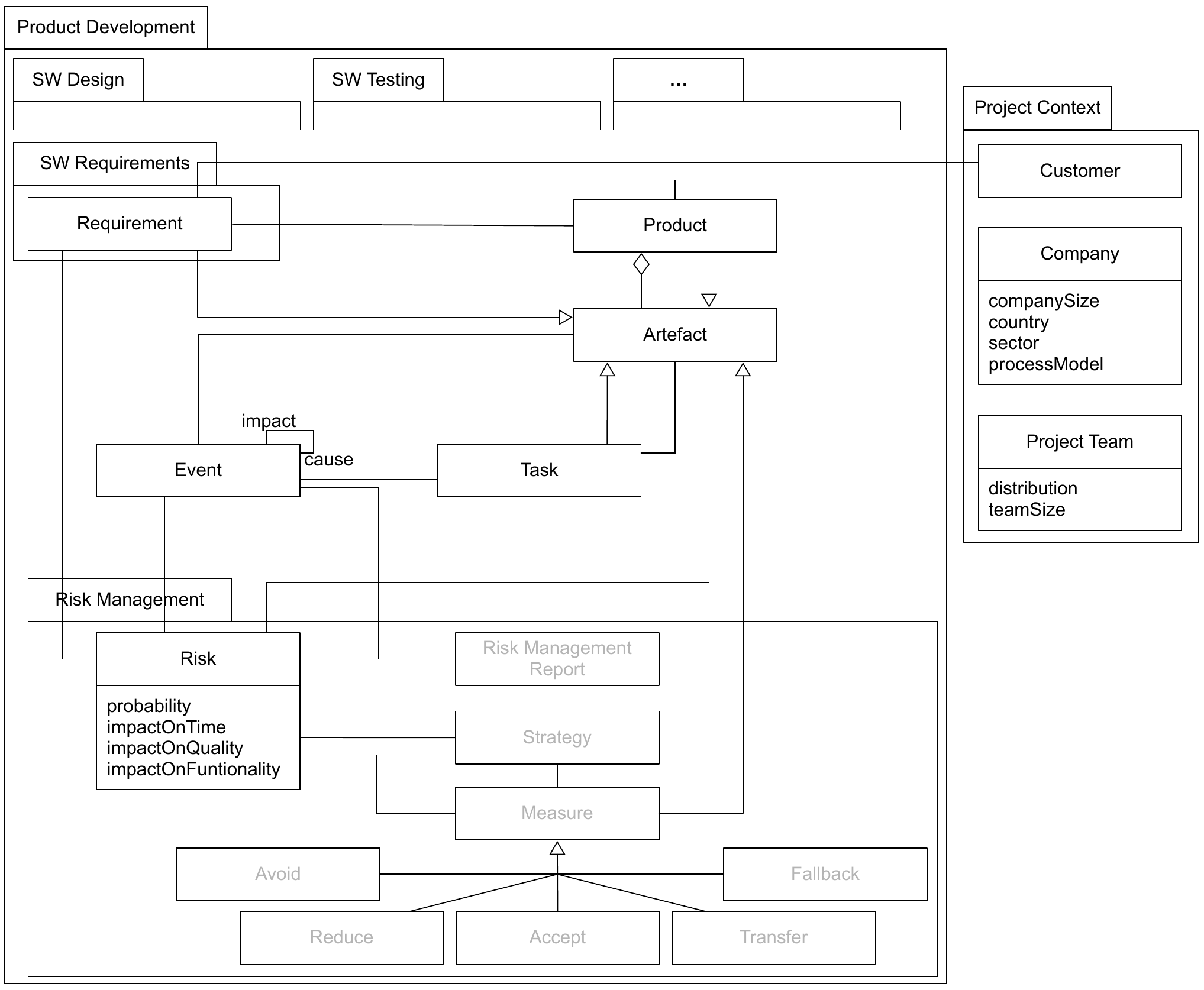}\\
  \caption{Strongly simplified meta model for evidence-based RE risk management. Darker elements encompass entities relevant for risk \emph{assessment which} is in scope of first development iteration. Elements in light grey encompass entities necessary for extension to a holistic risk \emph{management} approach.}
  \label{fig:MetaModel}
\end{figure}

To this end, we conceptualise the elements and constructs necessary for RE risk management via a meta model. This meta model captures the elements necessary for a holistic risk management in RE. Figure~\ref{fig:MetaModel} visualises a simplified resulting model, while highlighting those elements in scope of the initial approach for a risk assessment, i.e. the approach used for our first iteration presented in this paper. In centre of our attention are events. Events are phenomena that materialise in a measurable manner as well as their causes and effects. Those events either impact tasks carried out in a project or artefacts created / modified in the course of such tasks and can occur with various impacts at a certain probability (i.e. risks). 

Of further interest are the context factors used to characterise the development project, such as team characteristics or company characteristics. Please note that the meta model is intentionally incomplete and simplified and shall serve the sole purpose of highlighting those elements in scope of risk assessment. Out of scope, yet necessary for a holistic risk management are the implementation of measures used to, e.g., avoid certain risks.

To apply our approach in practical settings, we implement it via a light-weight prototype. To this end, we rely on spreadsheets realised via MS Excel for the presentation and reporting of risk factors and interactive calculation of the probabilities to encounter those factors based on a current situation (as learnt based on the data analysis described above). We further implement a graphical visualisation of all RE phenomena to increase the awareness of a current project situation using NodeXL. Details on the outcome can be taken from Sect.~\ref{sec:EbRE} where we introduce the resulting approach. 

\subsection{Validation}
\label{sec:ValidationDesign}
While the prototypical implementation already serves as an initial proof of concept, we want to better understand the suitability of the approach to practical contexts. This helps us preparing for the large-scale evaluation depicted at the upper part of Fig.~\ref{fig:ResearchDesign}. To this end, we conduct structured interviews with practitioners in face to face meetings using their answers to potentially revise the proposed (candidate) approach for a broader evaluation. In those interviews, we use the instrument summarised in Tab.~\ref{tab:EQuestionnaire} and developed from scratch in a curiosity-driven manner.
\begin{table}[htb]
\scriptsize
\centering
\caption{Summary of the interview questions (structure: M=metadata, A=approach, T=tool; type: FT=open free text, SC=closed single choice, LI=Likert scale rating).} 
\label{tab:EQuestionnaire} 
\begin{tabular}{clp{10.4cm}c}
\toprule  
& \textbf{ID} & \textbf{Question} & \textbf{Type} \\ 
\midrule
	M & 1 & What is the size of your company? & FT\\
	M & 2 & Please briefly describe the main sector of your company and the application domain of the software you build. & FT  \\ 
	M & 3 & How would you characterise the projects that you typically work in? & FT  \\ 
	M & 4 & What is your role in those projects? & FT  \\ 
	M & 5a & Is there already a methodology to manage risks in your company? & SC\\ 
	M & 5b & How do you manage risks in your company? & FT\\ 
	M & 6 & Which characteristic values would you consider for risk assessment? &  FT \\ 
	M & 7 & Does this risk management also address risks in Requirements Engineering? &  FT \\ 
	M & 8 & How important is it to address RE Risks in Risk Management? &  LI  \\ 
	M & 9 & Do you think it makes sense to distinguish RE Risk Management from the overall Risk Management?  &  LI  \\ 
	\midrule
	A & 10 & Do you think the approach is more precise than other risk assessment approaches?  &  SC \\ 
	A & 11a &  Do you think it makes sense to assess the risks based on phenomena? & SC  \\ 
	A & 11b & Why?  &  FT \\ 
	A & 12 & What are the benefits the approach?  &  FT \\ 
	A & 13 & What do you dislike about the approach?  &  FT \\ 
	A & 14 &  What are the barriers for the approach? &  FT \\ 
	A & 15 &  What should we do to improve the approach?  & FT  \\ 
	\midrule
	T & 16 & How appropriate do you think of criticality as a parameter for risk assessment?  & LI  \\ 
	T & 17 & Do you think using the tool would improve your performance in doing your job?  & LI  \\ 
	T & 18 & Do you think using the tool at work would improve your productivity? &  LI \\ 
	T & 19 & Do you think using the tool would enhance your effectiveness in your job? & LI  \\ 
	T & 20 & Would you find the tool useful in your job? & LI  \\ 
	T & 21 & Do you think learning to operate the tool would be easy for you?  & LI  \\ 
	T & 22 & Would you find it easy to get the tool to do what you want to do? &  LI \\ 
	T & 23 & Do you think it would be easy for you to become skilful in the use of the tool?  & LI  \\ 
	T & 24 & Do you find the tool easy to use? & LI  \\ 
	T & 25a & Would you use such a tool for RE risk assessment?  &  LI \\ 
	T & 25b & Why?  & FT  \\ 
	T & 26 & What would be necessary for using the tool?  & LI  \\ 
	T & 27a & How would you rate the tool in total?  &  LI \\ 
	T & 27b & Why?  & FT  \\ 
	T & 28 & How would you rate the risk assessments provided by the tool in total?   & LI  \\ 
	T & 29 & Would you rather adopt a pre-implemented tool or implement it with the risks that applied in your company?   & SC  \\ 
	 & 30 & Do you have further remarks?  &  FT \\ 
	\bottomrule
\end{tabular}
\end{table}
Further details on the validation can be taken from Sect.~\ref{sec:validation}.

\subsection{Large-scale Evaluation and Next Steps}
Once we understand the practitioners' perceptions of our light-weight approach allowing for last modifications of our risk assessment approach, we can scale up to practice by applying it in a longitudinal field study. Based on the outcome of the large-scale evaluation, we intend to gather knowledge about further barriers and limitations in context of risk management in RE. Further, such an evaluation will allow us to further scaling up by integrating proper measures for mitigating potential risks, thus, it allows us for a problem-driven (methodological) transition from risk assessment to risk management. These last steps are in scope of current activities at the time of writing this paper. 

\subsection{Validity Procedures}
It is noteworthy that the proposed approach relies on some assumptions on the NaPiRE data, which could represent threats to validity. The most important one concerns the accuracy of the coding process used to analyse the answers to the open questions RE problems, causes, and effects (used as a basis for the described risk assessment quantification). Coding is essentially a creative task with subjective facets of coders like experience, expertise and expectations. This threat was minimised by peer-reviewing the coding process. Concerning the NaPiRE survey itself, it was built on the basis of a theory induced from available studies and went through several validation cycles and pilot trials. Further details on the research methods employed can be taken from our previous publication~\cite{MW+16}

\section{Evidence-based RE Risk Assessment Approach}
\label{sec:EbRE}
This section presents the synthesised evidence-based risk management approach. Section~\ref{sec:OVerallConcept} introduces into the overall approach from a conceptual level, and Sect.~\ref{sec:PrototypeImpl} presents the prototypical implementation.

\subsection{Overall Approach to evidence-based RE Risk Assessment}
\label{sec:OVerallConcept}

\begin{figure}[!hbtp]
\centering
  \includegraphics[width=\textwidth]{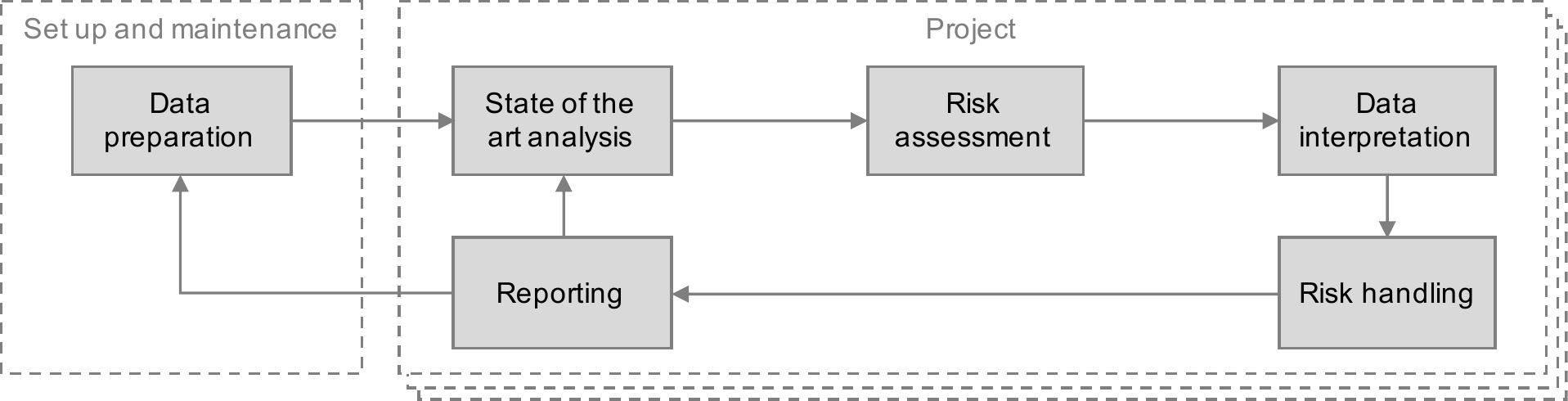}\\
  \caption{Elementary concepts of the RE Risk Assessment approach.}
  \label{fig:RMApproach}
\end{figure}

The risk assessment is integrated with an overall project-based risk management. Figure~\ref{fig:RMApproach} provides the approach's concept. Right in a \emph{Project}, we integrate the RE risk assessment with the overall project's risk management and expect a continuous implementation, i.e. we assume that a company implements its own risk management as a continuous task performed throughout the whole project.

Key to the approach presented is the data used for the (initial) risk assessment (Fig.~\ref{fig:RMApproach}; left). The required data can either taken from an external data pool, in our case NaPiRE data (see Sect.~\ref{sec:researchmethodology}), or from historical company data. Eventually, the company has to decide whether or not grounding the risk assessment in internal or external data. Using internal data brings in the opportunity to learn from past projects, i.e. to improve the company-specific risk management approach based on lessons learnt. Using external data instead, however, provides advantages, too. We postulate that by using cross-company data, risks that might have not materialised so far in own project settings can be captured based on third-party experience. Once the data source(s) has been selected, data needs to be prepared in order to carry out the risk assessment. Specifically, cause-effect models need to be developed (e.g., using Netica; Fig.~\ref{fig:Netica}) to provide processable input for the analysis tool (Sect.~\ref{sec:PrototypeImpl}). In a project, five tasks are carried out: 
\begin{enumerate}
	\item In a state-of-the-art analysis, the current project status is evaluated, possible risks, and already materialising risks are analysed---usually by looking for phenomena that ``announce'' an upcoming problem.
	\item In the assessment, based on the input data, the phenomena and associated risk (cause and effects) are calculated and the impact is estimated.
	\item In the interpretation, the risks are analysed and prioritised based on the computed probabilities and expected impact (e.g., using a risk classification matrix as used in PRINCE2 or a self-defined criticality index; see Sect.~\ref{sec:PrototypeImpl}).
	\item Based on the prioritisation, defined actions to adequately respond to the risks identified are initiated.
	\item Eventually, the analysis results and the actions initiated are reported. Reporting is used to complete the cycle and prepare the next assessment iteration, i.e. updated reporting data complement the general data basis. Also, reporting helps to evolve the general data basis, such that further projects can benefit from the reported findings.
\end{enumerate}

\subsection{Prototypical Implementation}
\label{sec:PrototypeImpl}

\begin{figure}[!hbtp]
\centering
  \includegraphics[width=\textwidth]{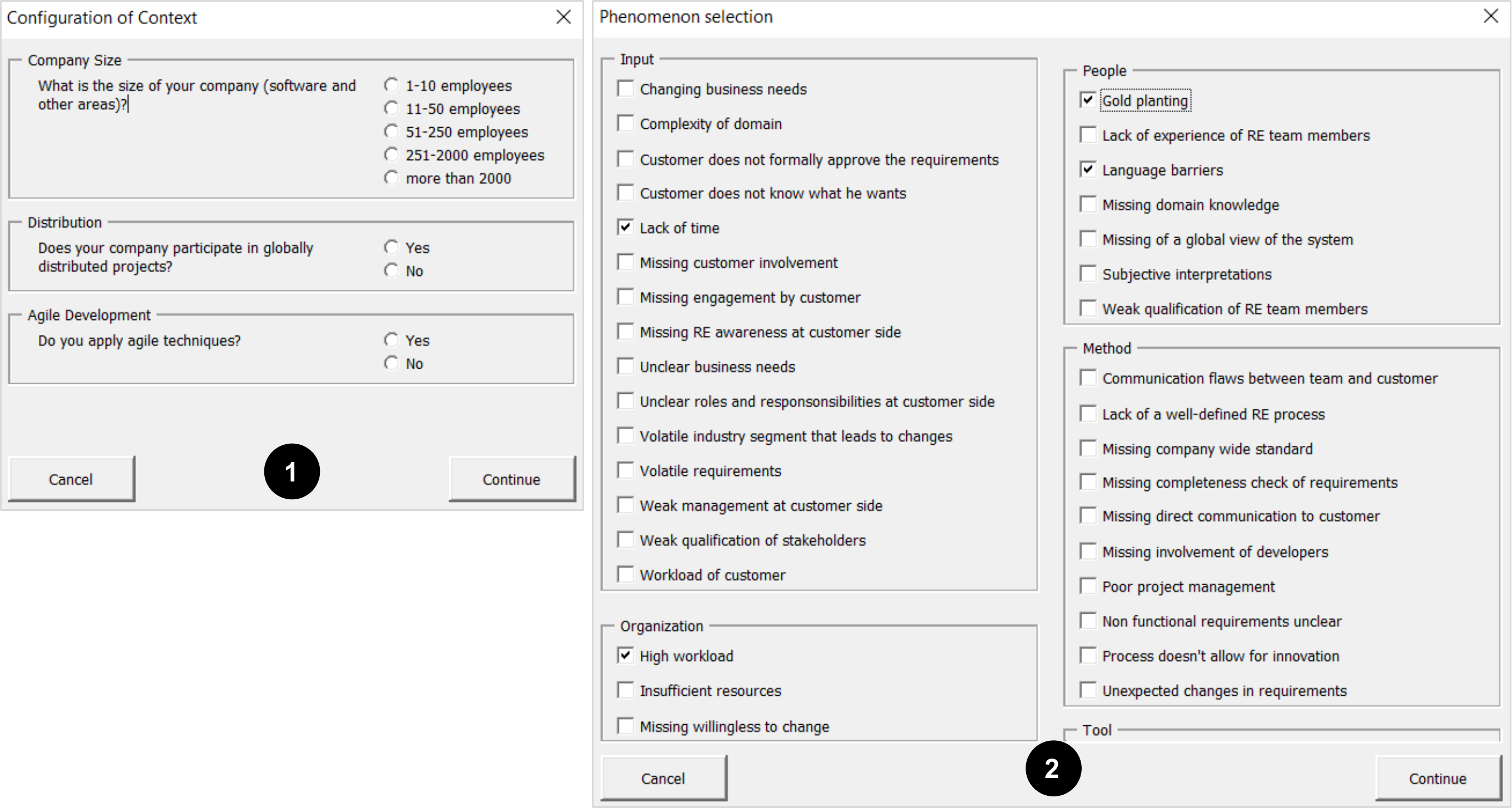}\\
  \caption{Exemplary configuration of the risk assessment tool (basis: NaPiRE data).}
  \label{fig:ToolConfig}
\end{figure}

The overall approach was developed using a variety of tools. Besides the tools used for the analysis, i.e. the modelling of the dependency networks and the probability graphs (Fig.~\ref{fig:Netica}, Netica), a prototypical support tool was realised using Microsoft Excel. The tool is applied to a given project setup in two steps. Figure~\ref{fig:ToolConfig} shows the first step---the configuration. In the configuration, the user is asked to characterise the project of interest in terms of the context factors company size, distributed work, and use of agile software development. The second part of the configuration aims at selecting all phenomena (see Sect.~\ref{sec:ProblemAnalysis}) of interest.

\begin{figure}[!hbtp]
\centering
  \includegraphics[width=\textwidth]{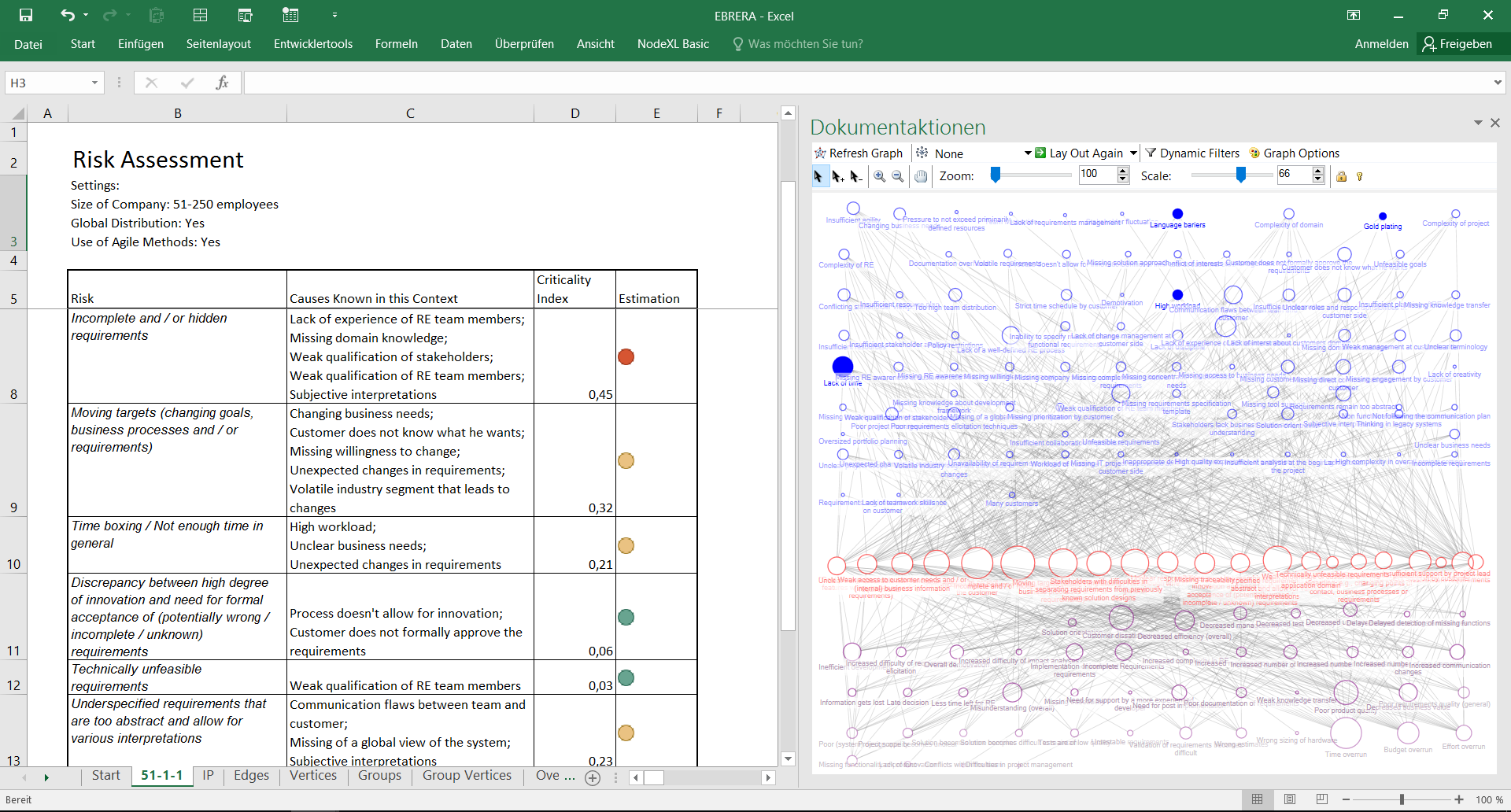}\\
  \caption{Example of a risk assessment report based on the the configuration from Fig.~\ref{fig:ToolConfig}.}
  \label{fig:ToolResult}
\end{figure}

Based on the selection, in the second step, the tool generates a report (Fig.~\ref{fig:ToolResult}) that comprises two parts: first, a risk assessment report is generated, which includes the risks and the known context-related causes associated with the phenomena chosen. The assessment report further present the criticality index and an estimate visualising the likelihood of a risk materialising in the current context. In the prototypical implementation, we use a custom criticality index, which is calculated as follows:
\[
	\frac{p_{i}}{n}\cdot\frac{p_{ij}}{n_{j}}\cdot\left(1+\frac{c_{i_{\tiny\emph{TRUE}}}}{c_{i}}\right)
\]
The index is calculated with the sum of the weighted phenomena $c_{i}$, the sum of the weighted phenomena $c_{i_{\tiny\emph{TRUE}}}$ that apply in the current context, the size $n$ of the data set used and the subset $n_{j}$ used for the phenomena that apply, the frequency of an identified problem in the whole data set $p_{i}$, and the frequency of a problem $p_{ij}$ in the subset under consideration.

The second part of the report visualises the the cause-effect graph in which the selected phenomena are highlighted and embedded in the overall network. This allows for inspecting the phenomena and finding orientation in the complex model.

\section{Validation}
\label{sec:validation}

Our overall objective in our validation is to better understand the current practitioners' views and acceptance of our solution proposal and get first feedback on potential barriers and limitations they see in the presented approach and the proposed tool. This shall help us in preparing for the large-scale evaluation in practice and the dissemination of our solution proposal. To this end, we conduct interviews with 6 practitioners from different companies and formulate the research questions listed in Tab.~\ref{tab:RQs}. 

\begin{table}[htbp]
\centering
\begin{tabular}{lp{10cm}}
\toprule
\bf{RQ~1}&How are requirements engineering risks assessed in projects?\\
\bf{RQ~2}&What are the benefits and drawbacks of the presented approach?\\
\bf{RQ~3}&What are the barriers for using the presented approach?\\
\bf{RQ~4}&Could practitioners imagine to use the presented tool in practice?\\
\bf{RQ~5}&What changes would be necessary to employ it in project settings?\\
\bottomrule
\end{tabular}
\caption{Research questions for the interview study.}
\label{tab:RQs}
\end{table}
After exploring what the state of practice in risk assessment is in the respective project environments, we are first interested in the general perception of the benefits and drawbacks in the presented approach. To actively prepare for the evaluation and give us the chance for last improvements, we further want to know barriers practitioner see in using our approach, whether they can imagine using the tool, and what changes would be necessary for using our solution proposal. As a preparation for the questions on the tool (RQ~4 and RQ~5), we simulated a given scenario to show the respondents the functionality of the tool in a live demo. The detailed instrument used for the interviews is shown in Sect.~\ref{sec:ValidationDesign}.

Please note that for reasons of confidentiality, but also because the answers were all recorded in German, we have to refrain from disclosing the raw data. In the following, we briefly introduce into the demographics before addressing each of the presented research questions.

\subsection{Demographics}

Our six respondents represent companies which differ in the sector and constraints attached, thus, covering a broader demographical spectrum. All six respondents are situated in Germany and their contextual information are summarised in Tab.~\ref{tab:ContextInformation}.

\begin{table}[htbp]
\centering
\scriptsize
\begin{tabular}  {p{0,5cm}p{1,5cm}p{2cm}p{2cm}p{2cm}p{2cm}p{2cm}}
\toprule
\textbf{ID} &\textbf{Company size}&\textbf{Main company sector}&\textbf{Role in projects}&\textbf{Project size}&\textbf{Process model}&\textbf{Project distribution}\\
\midrule
	1&300&IT consulting&Business analyst&4-10& Scrum& Nationally distributed\\ 
	2&100&Quality management, testing&External consultant for RE and Testing&Medium and large&Agile: Scrum, Kanban& Internationally distributed\\ 
	3&6000-6500&Software development and consulting&Internal consultant&20-40&Scrum, RUP&Internationally distributed\\ 
	4&20000&Accounting, IT-consulting&IT (Risk)-consultant&N/A&N/A&N/A\\ 
	5&30000&Insurance&Project manager&2-120&Agile and traditional approaches&Mostly national distributed\\ 
	6&180000&Software consulting, customer solution development&Requirements engineer &35&Waterfall, agile approaches&Internationally distributed\\
\bottomrule
\end{tabular}
\caption{Respondents overview (anonymised).}
\label{tab:ContextInformation}
\end{table}

In the following, we introduce the interview results. Where reasonable and possible, we provide original statements. Please note, however, that these statements are translated from Germany and in parts slightly paraphrased for the sake of understandability. 

\subsection{RQ~1: State of Practice in RE Risk Assessment}
The state of practice in RE risk assessment is as diverse as the respondents' environments. Some have already defined a risk management approach, others are currently establishing one. The most mature risk assessment approach we encountered was by respondent 3 who reports that at the beginning of each project, a cost estimation is conducted that includes the assessment of the relative risk of the project by means of a predefined checklist. 

As for the characteristics valued most important in context of risk assessment (Q~8), following statement seems most representative for the respondents:

\begin{center}
\setlength\fboxsep{0pt}
  \setlength\fboxrule{0pt}
\colorbox{lightgray}{
\fbox{\parbox{0.9\columnwidth}{
\begin{center}
\emph{``Probability of the risks and the potential monetary impact (including damages to the reputation)''}
\end{center}
}}}
\end{center}

When asked if their current risk assessment approach also addresses risks in RE (Q~7), only 2 respondents stated that requirements-related aspects were considered.

As for the importance to address RE risks in Risk Management (Q~8), the respondents had overall an agreement attitude: 

\begin{center}
\setlength\fboxsep{0pt}
  \setlength\fboxrule{0pt}
\colorbox{lightgray}{
\fbox{\parbox{0.9\columnwidth}{
\begin{center}
\emph{``[...] the sooner [risk management] starts, the better'' as it enables “to identify problems at the beginning.”}
\end{center}
}}}
\end{center}
On a scale from one to five, the respondents from two companies rate it with the highest importance, three with four and only one makes the rating dependent from the contract type. In case of a fixed price contract, he rates it with 4-5; for time and material contracts, he rates the importance with 1-2 only.

\subsection{RQ~2: Benefits and Drawbacks of the Approach}

When asked on the benefits and drawbacks of the presented approach, the answers provided a rich and diverse picture. 3 of 6 respondents clearly stated that the presented approach would be more precise for RE than available approaches (Q~10) while only 1 stated it would be not. Further, 4 of 6 respondents stated it would make sense to assess risks based on the empirically grounded RE phenomena (Q~11a), while, again, only 1 did not agree. When asked for their rationale (Q~11b), most respondents stated that such an evidence-based view would increase the awareness for risks in RE.  A further positive aspect highlighted was that the approach would rely on previously made experiences in same or similar environments:

\begin{center}
\setlength\fboxsep{0pt}
  \setlength\fboxrule{0pt}
\colorbox{lightgray}{
\fbox{\parbox{0.9\columnwidth}{
\begin{center}
\emph{``Embarrassing to repeat the same mistakes.''}
\end{center}
}}}
\end{center}

Challenges associated with the approach included: 

\begin{center}
\setlength\fboxsep{0pt}
  \setlength\fboxrule{0pt}
\colorbox{lightgray}{
\fbox{\parbox{0.9\columnwidth}{
\begin{center}
\emph{``It makes sense, but it seems difficult to realise in our company'' as well as ``Projects are not very comparable and, thus, we would need larger amounts of data. Sizing the clusters [sets of context factors] is challenging.''}
\end{center}
}}}
\end{center}

When asked directly for the benefits of the approach (Q~12), our respondents stated that it would offer an experience-based continuous learning including risks one might not be fully aware of, and it would offer the possibility to compare different project situations based on common risk factors. This would improve the situation for project leads by gaining more control.

When asked for what they disliked (Q~13), respondents stated, in one form or the other, that transparency on risk factors experienced by other projects with similar characteristics might lead to more insecurity of the project team:

\begin{center}
\setlength\fboxsep{0pt}
  \setlength\fboxrule{0pt}
\colorbox{lightgray}{
\fbox{\parbox{0.9\columnwidth}{
\begin{center}
\emph{``Too much data may lead to confusion.''}
\end{center}
}}}
\end{center}

Our respondents further stated that the maintainability of such an approach would constitute another challenge:
\begin{center}
\setlength\fboxsep{0pt}
  \setlength\fboxrule{0pt}
\colorbox{lightgray}{
\fbox{\parbox{0.9\columnwidth}{
\begin{center}
\emph{``Keeping the data up-to-date is an elaborate task [as we need to] ensure high data quality.''}
\end{center}
}}}
\end{center}

\subsection{RQ~3: Barriers for using the Approach}

As the main barriers for using the presented approach (Q~14), our respondents stated, in tune with the previous statement, the need to maintain large amounts of data. A further statement included that a strong focus on risks related to RE might lead to pessimism. A further perceived barrier affected the fear for projects giving up their individual characteristics when being compared with other projects:

\begin{center}
\setlength\fboxsep{0pt}
  \setlength\fboxrule{0pt}
\colorbox{lightgray}{
\fbox{\parbox{0.9\columnwidth}{
\begin{center}
\emph{``Single projects would have the impression their individual characteristics would be neglected.''}
\end{center}
}}}
\end{center}

When asked what we could do to improve the approach (Q~15), we encountered mainly two aspects: Inclusion of time as a dimension into the phenomena that would allow to calculate the time until risks may manifests as real events, and a better visualisation for especially large amounts of data.

\subsection{RQ~4: Usability of the Tool}
\begin{figure}[!hbtp]
\centering
  \includegraphics[width=\textwidth]{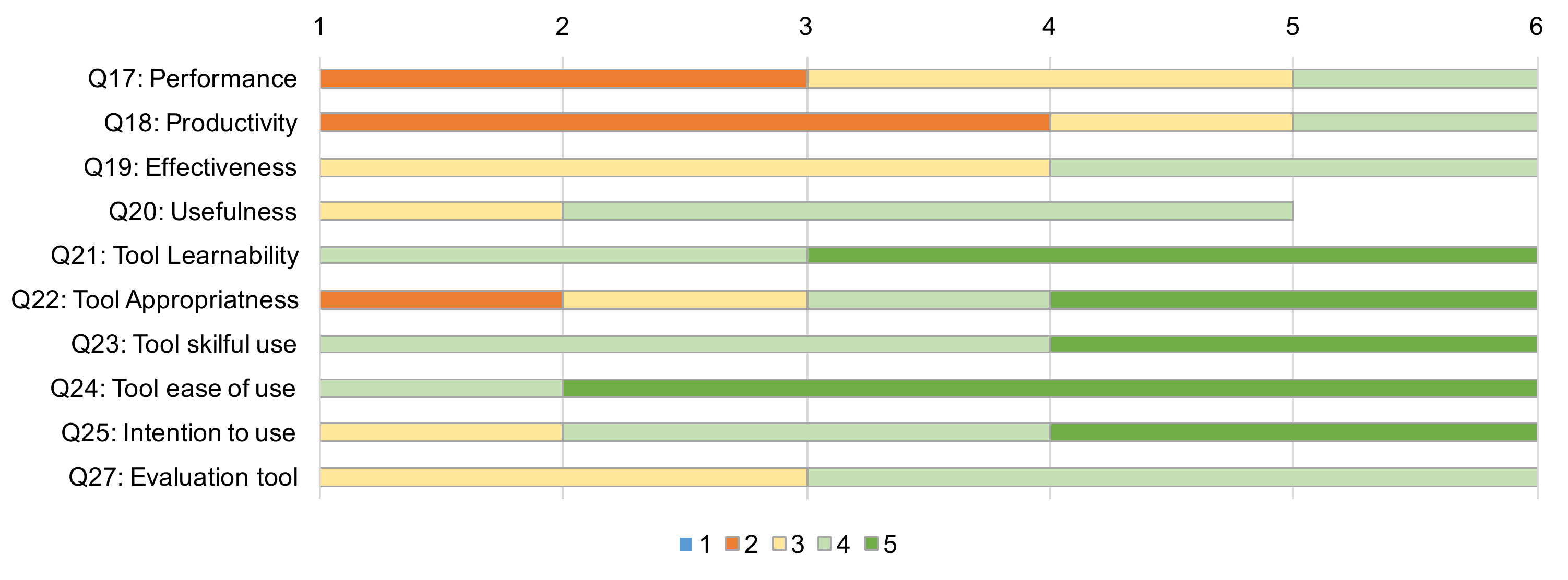}\\
  \caption{Evaluation of the concept and the tool prototype with the six practitioners. The figure summarises the number of responses as stacked bar charts, ordered by their agreement on Likert scales from 1 (low/no agreement) to 5 (high agreement).}
  \label{fig:StatsEvaluation}
\end{figure}

As for the usability of the tool presented via a live demo, we asked our respondents the questions from Tab.~\ref{tab:EQuestionnaire} and visualise the results in Fig.~\ref{fig:StatsEvaluation}. While the focus of our validation was on qualitative feedback, it is still noteworthy that the prototypical implementation yielded overall a reasonably good acceptance while acknowledging that---although the tool offers a light-weight mechanism to assess the risks---it still requires effort for the data maintenance and curation (reflected in a mixed rating in the performance and productivity).

\subsection{RQ~5: Necessary Changes}

When asked for the necessary changes we would need to apply to the whole approach and to the corresponding prototypical tool (Q~26), two major aspects were prominent:
\begin{center}
\setlength\fboxsep{0pt}
  \setlength\fboxrule{0pt}
\colorbox{lightgray}{
\fbox{\parbox{0.9\columnwidth}{
\begin{center}
\emph{``Support in the data collection and [possibilities to better] compare company data with the data of other companies'' as well as ``Improved reliability of the data''.}
\end{center}
}}}
\end{center}
Those statements are, in fact, in tune with our expectations from the current stage of development were we could only use the context factors known from the baseline data in NaPiRE. Overall, our respondents concluded along that line, which can be summarised by this exemplary statement:
\begin{center}
\setlength\fboxsep{0pt}
  \setlength\fboxrule{0pt}
\colorbox{lightgray}{
\fbox{\parbox{0.9\columnwidth}{
\begin{center}
\emph{``Detailed project characteristics need to be included [as context factors]''}
\end{center}
}}}
\end{center}
as well as by the need to further scale up from risk assessment to a holistic risk management as exemplarily reflected in this statement:
\begin{center}
\setlength\fboxsep{0pt}
  \setlength\fboxrule{0pt}
\colorbox{lightgray}{
\fbox{\parbox{0.9\columnwidth}{
\begin{center}
\emph{``Countermeasures should be included''}
\end{center}
}}}
\end{center}

In summary, the need to increase the precision and reliability of the data by adding more data points as well as context factors, and the need to extend the approach with context-sensitive recommendations in form of guidelines is in tune with our hopes and expectations expressed in Sect.~\ref{sec:researchmethodology}. This strengthens our confidence in the suitability to release our approach for a large-scale evaluation which is in scope of current activities.

\section{Conclusion}
\label{sec:conclusion}

In this paper, we elaborated on the need for an evidence-based risk management approach that considers the particularities of RE. We argued that for such an approach, we would need a proper empirical basis on practical problems, root causes and effects in industrial RE. To lay a first foundation for such a risk management approach, we presented a cyclic research design that makes use of empirical data gathered from the NaPiRE project yielding a first approach for an evidence-based RE risk assessment. We implemented this risk assessment approach and conducted a first validation by interviewing 6 respondents from different companies. 

The results of the validation have shown potential for improvement, but it also strengthens our confidence in the maturity of our solution proposal and its suitability to be transferred into practice and applied in a large-scale evaluation. In particular, the feedback from the practitioners indicate an increase of awareness for individual risk factors in RE which so far could not be provided by existing traditional risk management approaches. They also strengthened our confidence in continuing the envisioned future work. This work includes running a field study which is currently in preparation and enriching our approach with a set of recommendations associated with context-sensitive risk factors as shown in previous work on guidelines to prevent RE problems~\cite{kalinowski:swqd16, mafra2016}.

\paragraph{Acknowledgements}
We are grateful to all practitioners who participated in the evaluation and who shared their experiences and insights into their environments.

\bibliographystyle{splncs}
\bibliography{Literature} 
\end{document}